# Manipulation of Exciton Dynamics in Single-Layer WSe$_2$ Using a Toroidal Dielectric Metasurface

Long Yuan[1#*], Jeeyoon Jeong[2,3#], Kevin Wen Chi Kwock[4], Emanuil S. Yanev[5], Michael Grandel[5], Daniel Rhodes[6], Ting S. Luk[2], P. James Schuck[5], Dmitry Yarotski[1], James C. Hone[5], Igal Brener[2*] and Rohit P. Prasankumar[1*]

**Affiliations:**

[1]Center for Integrated Nanotechnologies, Los Alamos National Laboratory, Los Alamos, New Mexico 87545, United States

[2]Center for Integrated Nanotechnologies, Sandia National Laboratories, Albuquerque, New Mexico 87185, United States

[3]Department of Physics and Institute for Accelerator Science, Kangwon National University, 1 Gangwondaehak-gil, Chuncheon-si 24341, Gangwon-do, Korea

[4]Department of Electrical Engineering, Columbia University, New York, New York, 10027, United States

[5]Department of Mechanical Engineering, Columbia University, New York, New York, 10027, United States

[6]Department of Material Science and Engineering, University of Wisconsin-Madison, Madison, Wisconsin, 53706, United States

[#]These authors contributed equally to this work.

[*]ibrener@sandia.gov; rpprasan@lanl.gov; lyuan@lanl.gov

**ABSTRACT:** Recent advances in emerging atomically thin transition metal dichalcogenide semiconductors with strong light-matter interactions and tunable optical properties provide novel approaches for realizing new material functionalities. Coupling two-dimensional semiconductors with all-dielectric resonant nanostructures represents an especially attractive opportunity for manipulating optical properties in both the near-field and far-field regimes. Here, by integrating single-layer $WSe_2$ and titanium oxide ($TiO_2$) dielectric metasurfaces with toroidal resonances, we realized robust exciton emission enhancement over one order of magnitude at both room and low temperatures. Furthermore, we could control exciton dynamics and annihilation by using temperature to tailor the spectral overlap of excitonic and toroidal resonances, allowing us to selectively enhance the Purcell effect. Our results provide rich physical insight into the strong light-matter interactions in single-layer TMDs coupled with toroidal dielectric metasurfaces, with important implications for optoelectronics and photonics applications.

# INTRODUCTION

The burgeoning interest in two-dimensional (2D) transition metal dichalcogenides (TMDs) with tunable optical and electronic properties offers unprecedented opportunities for novel optoelectronics and photonics applications[1-4]. The large in-plane quantum confinement in single-layer TMDs and reduced dielectric screening leads to the formation of tightly bound excitons that exhibit strong light-matter interactions[5,6]. However, the relatively low quantum yield in many of these nanomaterials significantly limits their photonics applications[3,7]. Moreover, the strong many-body exciton interactions in single-layer TMDs give rise to efficient exciton-exciton annihilation that further hampers their quantum efficiency in optoelectronic applications[8-11].

To overcome the aforementioned limitations, various methods have been proposed to enhance light-matter interactions in single-layer TMDs by incorporating robust near field electromagnetic enhancement. In this context, resonant dielectric metasurfaces (MS) fabricated from high refractive index materials have emerged as a powerful new approach for nanoscale manipulation of light[12-17]. The low absorption of dielectric metasurfaces at their resonant frequency reduces their losses, providing a major advantage over the more commonly used metallic plasmonic nanostructures[16,17]. These negligible losses of dielectric metasurfaces were instrumental in previous studies of emission control via quantum emitters coupled to metasurfaces[18]. Integrating single-layer TMDs with resonant metasurfaces can therefore provide an effective approach for tailoring light-matter interactions[19,20]. Previous demonstrations of TMD/MS hybrid structures have showcased enhanced optical properties such as absorption, directional emission and spontaneous emission rate[15,21,22]; however, to the best of our knowledge, the influence of dielectric metasurfaces on exciton dynamics in TMDs has never been studied and is key to many of the previously mentioned applications.

In this work, we integrated a cuboid $TiO_2$ metasurface with single-layer $WSe_2$ (1L-$WSe_2$) to demonstrate enhanced control of exciton emission and dynamics. The cuboid design not only provides a robust way to achieve a high quality-factor (Q) toroidal resonance, but also allows the enhanced electromagnetic field of the toroidal mode to leak outside of the dielectric resonators[16], thereby enabling strong light-matter interactions with the excitons in 1L-$WSe_2$. We observed a robust photoluminescence (PL) enhancement of over one order of magnitude for the 1L-$WSe_2$/MS heterostructure at both room and low temperatures. The spectral overlap between the neutral exciton resonance and the toroidal resonance is maximized at 290 K and gradually decreases with decreasing temperature, while other localized excitonic states begin to spectrally overlap with the toroidal resonance. This enabled us to selectively enhance radiative rates and suppress exciton-exciton annihilation for specific excitonic states by temperature tuning them to spectrally overlap with the metasurface resonance. Our results demonstrate that a dielectric metasurface with a high-Q toroidal resonance can enable the efficient manipulation of exciton dynamics and annihilation in monolayer TMDs through enhanced light-matter interactions.

## RESULTS AND DISCUSSION

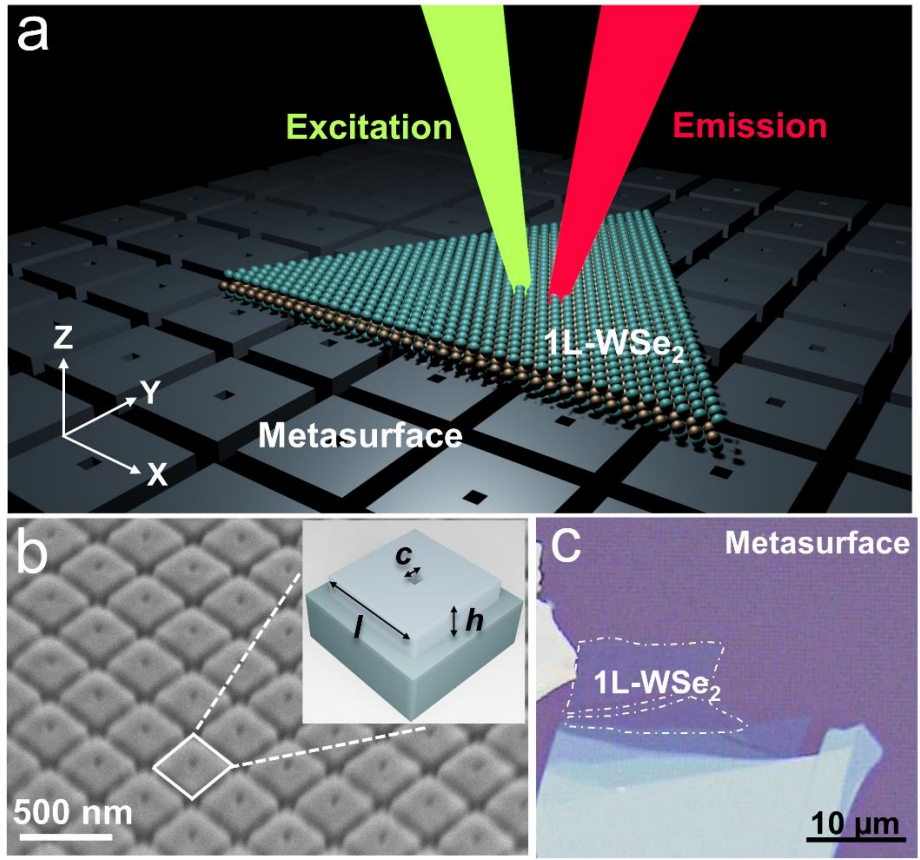

**Figure 1.** Integrating an atomically thin semiconductor with a cuboid dielectric metasurface. (a) Schematic of 1L-WSe$_2$ on top of a TiO$_2$ cuboid metasurface. The red and green beams represent PL emission from 1L-WSe$_2$/MS and the excitation laser beam, respectively. (b) SEM image of the dielectric metasurface; $l$, $h$, and $c$ represent the length, thickness, and void size of the metasurface, respectively. (c) Optical image of 1L-WSe$_2$ on the metasurface.

**Integrating 1L-WSe₂ with a high-Q toroidal dielectric metasurface and measuring the resulting PL enhancement.** Figure 1a displays a schematic of a 1L-WSe$_2$ film transferred on top of a cuboid TiO$_2$ metasurface. We first fabricated the metasurface sample using a standard top-down approach (more details in the Materials and Methods section). The sample consists of an array of TiO$_2$ resonators with side length $l$ = 450 nm, void size $c$ = 50 nm and film thickness $h$ = 100 nm, with periods of 500 nm in both directions. We then transferred the exfoliated 1L-WSe$_2$ sample on top of the metasurface using a standard dry-transfer method. Figure 1b displays a scanning electron microscope (SEM) image of the cuboid TiO$_2$ metasurface. An optical image of the 1L-WSe$_2$/MS hybrid structure is shown in Figure 1c, revealing good contact between them without any visible bubbles or contamination at the interface.

Figure 2a displays the reflectance spectrum of bare 1L-WSe$_2$ at 290 K, revealing the lowest-energy neutral exciton resonance at 1.67 eV, as well as a resonance at ~ 2.1 eV originating from direct band-to-band transitions at the K point and another at ~ 2.45 eV originating from band nesting[23,24]. The PL spectrum of 1L-WSe$_2$ exhibits a single sharp neutral exciton emission peak at 1.66 eV. The measured room temperature transmission spectrum of the metasurface (without 1L-WSe$_2$) displays sharp Fano-type resonances at 1.64 eV (T$_1$) and 1.69 eV (T$_2$), with Q factors of ~ 206 and 8, respectively (Figure S1). The Q factor should not substantially change at low temperatures, as TiO$_2$ has negligible absorption in the visible range at all measured temperatures, and other effects, like spectral shifts from thermal expansion and refractive index changes from thermo-optic effects, should be negligible for our structures.

We attribute T$_1$ to a toroidal dipole mode interfering with an equally strong magnetic quadrupole, and T$_2$ to a magnetic dipole mode[16]. Both T$_1$ and T$_2$ could spectrally overlap with the neutral exciton resonance in 1L-WSe$_2$ at room temperature. Due to the significantly larger Q factor of the T$_1$ resonance compared to the T$_2$ resonance, we expect that T$_1$ will have the most significant contribution to enhancing light-matter interactions. Figure 2b (top) displays the spatial distribution of the electric field for T$_1$ at a height of ~ 1 nm above the void area, simulated using linearly polarized excitation light and carried out using commercial finite-difference time-domain (FDTD) software. The polarization of the enhanced field predominantly follows that of the incident excitation. In addition, due to the isotropic in-plane dipole moment for excitons in monolayer WSe$_2$, the electric field distribution should not depend on the polarization or alignment of the crystal axes with the symmetry axes of the metasurface. We note that our metasurface resonances were designed to have an in-plane dipole mode for optimal overlap with WSe$_2$ excitons, especially since the electric field from out-of-plane dipole modes decays quickly in the out-of-plane direction, leading to a smaller field enhancement.

The simulated $(|E|/|E_0|)^2$ displays the highest enhancement (~ 300 fold) in the middle of the void area as shown in Figure 2b (top). Figure 2b (bottom) displays a cross-sectional view of the electric field distribution for T$_1$, revealing an enhancement of the electric field by a factor of ~ 700 fold in the center of the void area; it is not symmetric along the z-direction because of the substrate beneath the resonators. We also simulated the electric field distribution at a non-resonant condition (2.5 eV) and found no obvious enhancement (Figure S2).

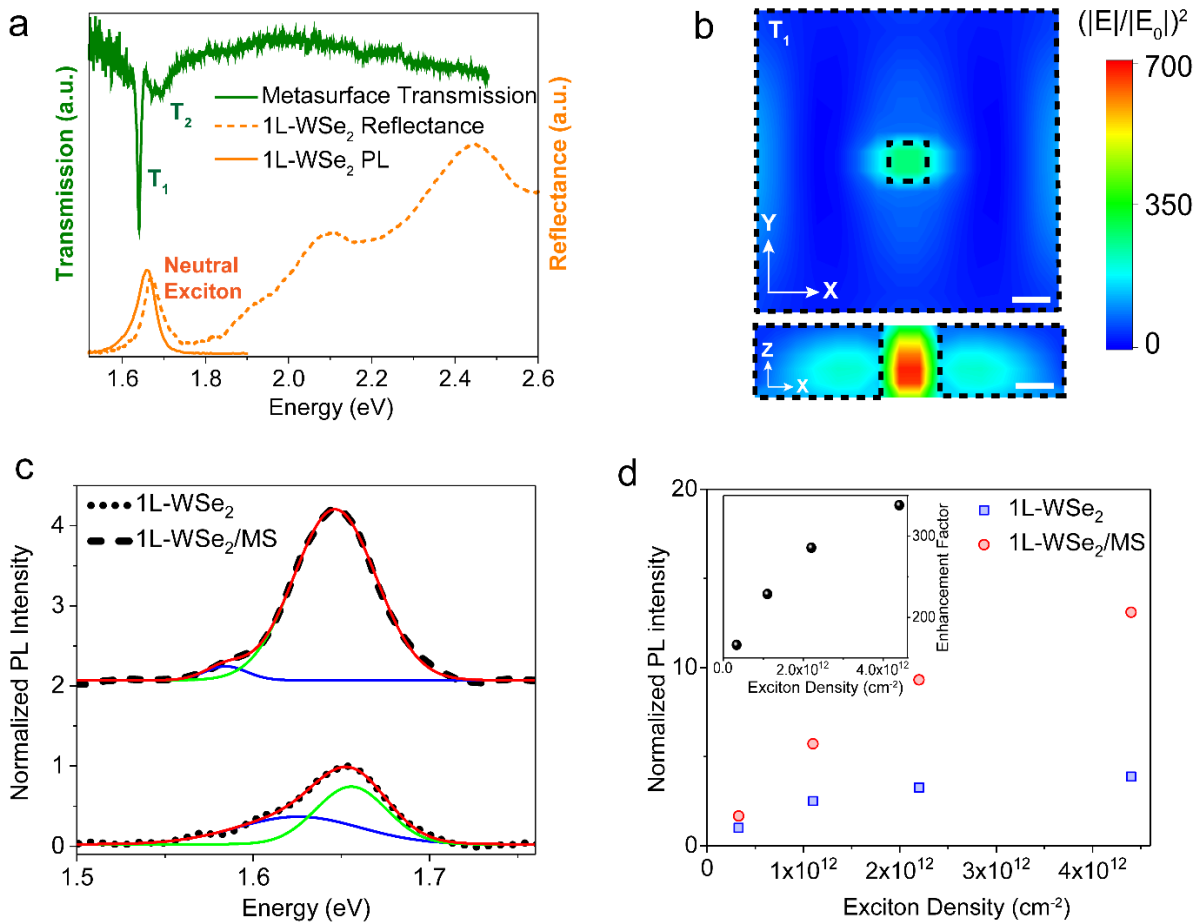

**Figure 2.** PL emission enhancement in 1L-WSe$_2$ coupled with a toroidal dielectric metasurface. (a) Measured transmission spectra of the metasurface (green line) as well as reflectance (orange dashed line) and PL spectra (orange solid line) of bare 1L-WSe$_2$ on quartz at 290 K. (b) Simulation of the electric field spatial distribution $(|E|/|E_0|)^2$ at a height of 1 nm above the central void (top); cross-sectional view of the electric field distribution (bottom). The electric field distribution was simulated at the toroidal resonance of 1.64 eV. The scale bar represents 50 nm, and the black dashed lines denote the unit cell of the metasurface.

(c) PL spectra measured on bare 1L-WSe$_2$ (bottom) and the 1L-WSe$_2$/MS heterostructure (top) at 290 K. The PL intensity was normalized by the peak emission intensity of bare 1L-WSe$_2$. The 1L-WSe$_2$/MS spectra was shifted vertically for clarity. The PL spectra were fitted (red solid lines) using two Gaussian functions (blue and green solid lines). The blue and green peaks are attributed to trion and neutral exciton emission, respectively (see main text). (d) PL emission from the 1L-WSe$_2$/MS heterostructure (circle) and bare 1L-WSe$_2$ (square) as a function of exciton density at 290 K. The inset displays the PL enhancement factor as a function of exciton density. The PL intensity was normalized by the emission intensity of bare 1L-WSe$_2$ at an exciton density of $3.3 \times 10^{11}$ cm$^{-2}$.

Figure 2c presents a quantitative comparison of PL spectra from bare 1L-WSe$_2$ and 1L-WSe$_2$/MS, measured on the same substrate. We define the effective PL enhancement factor as[15,25]:

$$Enhancement\ Factor = \frac{I_{on}}{A_{MS}} \cdot \frac{A_0}{I_{off}} \qquad (1)$$

where $I_{on}$ and $I_{off}$ are the PL intensity of 1L-WSe$_2$ with and without the metasurface, $A_0$ is the laser excitation area and $A_{MS}$ is the geometrical area of the center void where the field enhancement is the largest. The effective enhancement factor is determined to be ~ 170 at a low exciton density ($3.3 \times 10^{11}$ cm$^{-2}$) when normalized by the void area of the metasurface, but is comparable with previous measurements for monolayer MoS$_2$ coupled with dielectric nanodisks[19]. We note that the PL linewidth from 1L-WSe$_2$/MS is not substantially narrower than from 1L-WSe$_2$, since there are significant contributions to the PL signal from WSe$_2$ regions outside the nanoscale enhancement area. In addition, while the effect of the metasurface on exciton emission spectra is not particularly pronounced compared to prior studies coupling TMD monolayers with plasmonic or dielectric nanocavities[26-28], we still see that the dielectric metasurface selectively enhances PL emission when specific exciton states are temperature tuned to spectrally overlap with the toroidal resonance, modifying the emission spectra at both 290 K and 8 K. This is shown in Figure S3, where at 8 K, the metasurface resonance selectively enhances PL emission from the localized states, while there is no PL enhancement of the neutral exciton states due to the lack of spectral overlap. We also note that there is a small redshift in the exciton and trion peaks (~ 6 meV) in the 1L-WSe$_2$/MS compared with bare 1L-WSe$_2$, likely due to the change in dielectric environment[29].

More generally, PL enhancement in monolayer WSe$_2$ coupled with a dielectric metasurface can result from two contributions: enhancement of the incident photoexcitation and an enhanced spontaneous emission rate through the Purcell effect[15,30] showing that the exciton resonance in 1L-WSe$_2$ efficiently overlaps with the toroidal resonance of the dielectric metasurface. From our simulations (Figure S2), we found that the 1.64 eV toroidal resonance negligibly enhances the 2.48 eV photoexcitation, implying that the PL enhancement is mostly due to local density of states (LDOS) enhancement. Our simulations (Supporting Note 1) show a LDOS enhancement by a factor of ~ 5100 at the toroidal resonance when it fully overlaps with the exciton resonance. However, this defines the upper limit of PL enhancement, since the toroidal resonance only has a small field overlap with the exciton resonance in the real system. Moreover, our sample was placed at the surface, not at the middle of the voids where the field is strongest, which could also reduce the PL enhancement.

Additionally, we decomposed the PL spectra using a sum of two Gaussian functions originating from neutral exciton and trion (charged exciton) emission, respectively[31]. The emission intensity ratio between the neutral exciton and trion increased from 2:1 to 12:1 for bare 1L-WSe$_2$ and 1L-WSe$_2$/MS, respectively, due to the enhanced radiative recombination rate. Finally, we also measured the PL intensity as a function of exciton density for both bare 1L-WSe$_2$ and WSe$_2$/MS (Figure 2d). The PL intensity displays a sub-linear relationship with exciton density that can be attributed to exciton-exciton annihilation[32]. The corresponding PL enhancement factor (inset in Figure 2d) increases with exciton density due to the reduced exciton-exciton annihilation in the WSe$_2$/MS heterostructure. We will describe exciton-density-dependent measurements further in the following section. Overall, our PL studies underline our ability to tailor exciton emission in TMD monolayers via integrating with dielectric metasurfaces.

**Tailoring exciton dynamics through the Purcell effect by controlling the spectral overlap between excitonic and toroidal resonances.** As our 1L-WSe$_2$/MS heterostructure displayed significant PL enhancement through a strong Purcell effect, we investigated the expected increase in radiative recombination rates using ultrafast time-resolved PL spectroscopy. We used a streak camera with simultaneous spectral (~ 1 nm) and temporal (~ 5 picoseconds (ps)) resolution to resolve the PL dynamics for distinct exciton states in the 1L-WSe$_2$/MS heterostructure. Furthermore, by tuning the neutral exciton resonance energy in 1L-WSe$_2$ with temperature, we were able to tailor its spectral overlap with the toroidal resonance, giving an additional degree of control over exciton dynamics.

Figure 3a displays the temperature (T)-dependent PL spectra of bare 1L-WSe$_2$. The neutral exciton emission energy exhibits a well-defined blue shift as the temperature decreases, due to lattice contraction[33] as well as reduced electron-phonon interaction[34]. It is also clear that the spectral overlap between the neutral exciton resonance and the toroidal resonance is maximized at T = 290 K and gradually decreases with decreasing temperature. We then observed that the neutral exciton decays nearly twice as fast (340 ps) in the 1L-WSe$_2$/MS heterostructure than in bare WSe$_2$ (626 ps) at 290 K (Figure 3b), since the Purcell effect enhances the radiative recombination rate. In contrast, at T =150 K, the neutral exciton resonance has little spectral overlap with the metasurface resonance, causing the exciton dynamics in the heterostructure to be nearly identical to bare 1L-WSe$_2$. We note that the PL lifetime is only reduced by a factor of ~ 2, since the majority of the observed Purcell enhancement only occurs in the small cubic void area (~ 50 nm, Figure 2b). which is significantly smaller than the diffraction limited laser spot size (~ 400 nm) in our measurements. Furthermore, the Purcell factor is determined by both field overlap and cavity Q factor, which directly affect the coupling between the exciton and the metasurface resonance mode. This is one factor limiting the maximum lifetime reduction.

The timescales reported here agree well with previous studies of recombination dynamics in monolayer TMDs using time-resolved PL[7,9,31,32,35], although other studies with higher temporal resolution[36,37] have revealed coherent exciton dynamics due to homogeneous broadening on a timescale of hundreds of femtoseconds. The longer timescales measured here mostly reflect exciton recombination inhomogeneously broadened by defect/impurity, phonon or carrier scattering, due to the significantly lower temporal resolution of our experiments (~ 5 ps).

We also estimated the radiative and non-radiative recombination rates from our data (see Supporting Note 2 for more detail). For bare 1L-WSe$_2$, we obtained radiative and non-radiative rates of $4.9 \times 10^7$ and $1.5 \times 10^9$ s$^{-1}$, respectively, using a quantum yield of 0.03 from a previous study[35]. For the 1L-WSe$_2$/MS, the PL quantum yield is enhanced by a factor of two to 0.06, giving radiative and non-radiative rates of $1.7 \times 10^8$ and $2.8 \times 10^9$ s$^{-1}$, respectively. The radiative rate is thus enhanced by a factor of 3.5, significantly larger than the non-radiative rate enhancement of 1.9. This is expected, since the Purcell effect can increase the radiative rate, but cannot directly alter the non-radiative rate (which also limits the maximum lifetime reduction).

We then studied exciton emission and dynamics at T = 8 K (Figures 3c and 3d). Localized exciton states with a significantly lower emission energy than neutral excitons typically appear at low temperatures and are the primary contributor to the PL emission at T = 8 K[31] (Figure S4), where the neutral exciton emission intensity is significantly suppressed because the excitonic ground state for WSe$_2$ is optically dark[38]. The lower-energy emission may also originate from dark states or phonon replicas[39-41]. However, we ruled them out based on the following evidence. First, we deconvoluted the PL spectra of 1L-WSe$_2$ at 8 K using four Gaussian functions (Figure S4a).

The energy difference between localized state 1 and the neutral exciton is ~ 100 meV, two times larger than that between the dark exciton and the neutral exciton[39-41]. Also, the energy difference between localized states 2 and 3 is ~ 30 meV, significantly larger than that between the dark exciton and its phonon replica (~ 20 meV)[41]. Second, the PL decay of localized states, measured at a low exciton density ($3.3 \times 10^{11}$ cm$^{-2}$), was fit with a bi-exponential function with fast (55 ps) and slow (1.1 ns) decays (Figure S4b). However, dark exciton/phonon replicas display a single-exponential PL decay with a lifetime of ~ 200 ps[39-41]. Fast and slow components with similar timescales were also previously observed in localized state emission[42,43] and attributed to shallow trapped and deep trapped/defect bounded excitons, respectively, further supporting our interpretation.

Figure 3d plots the PL dynamics for the neutral exciton and a localized exciton peak that overlaps well with the toroidal resonance at T = 8 K. In contrast with the dynamics at T = 290 K, the localized exciton dynamics become faster in the 1L-WSe$_2$/MS (304 ps) than bare 1L-WSe$_2$ (415 ps) due to the Purcell effect-enhanced radiative rate, while the neutral exciton dynamics remain essentially the same in both 1L-WSe$_2$ and 1L-WSe$_2$/MS samples. We also carried out measurements at 30 K and observed a similar result (Figure S5). This further underlines the central role of spectral overlap between exciton and toroidal resonances in modulating exciton dynamics. In addition, we note that while previous studies have demonstrated enhanced PL emission and radiative recombination rates in single-layer TMDs coupled with plasmonic resonances, the observed enhancement occurs across all exciton states[25,44]. Here, the high-Q toroidal resonances enable us to selectively enhance emission and radiative rates by precisely tuning the toroidal resonance energy to couple with specific exciton resonances, demonstrating that dielectric metasurfaces can provide additional flexibility for manipulating light-matter interactions in TMDs.

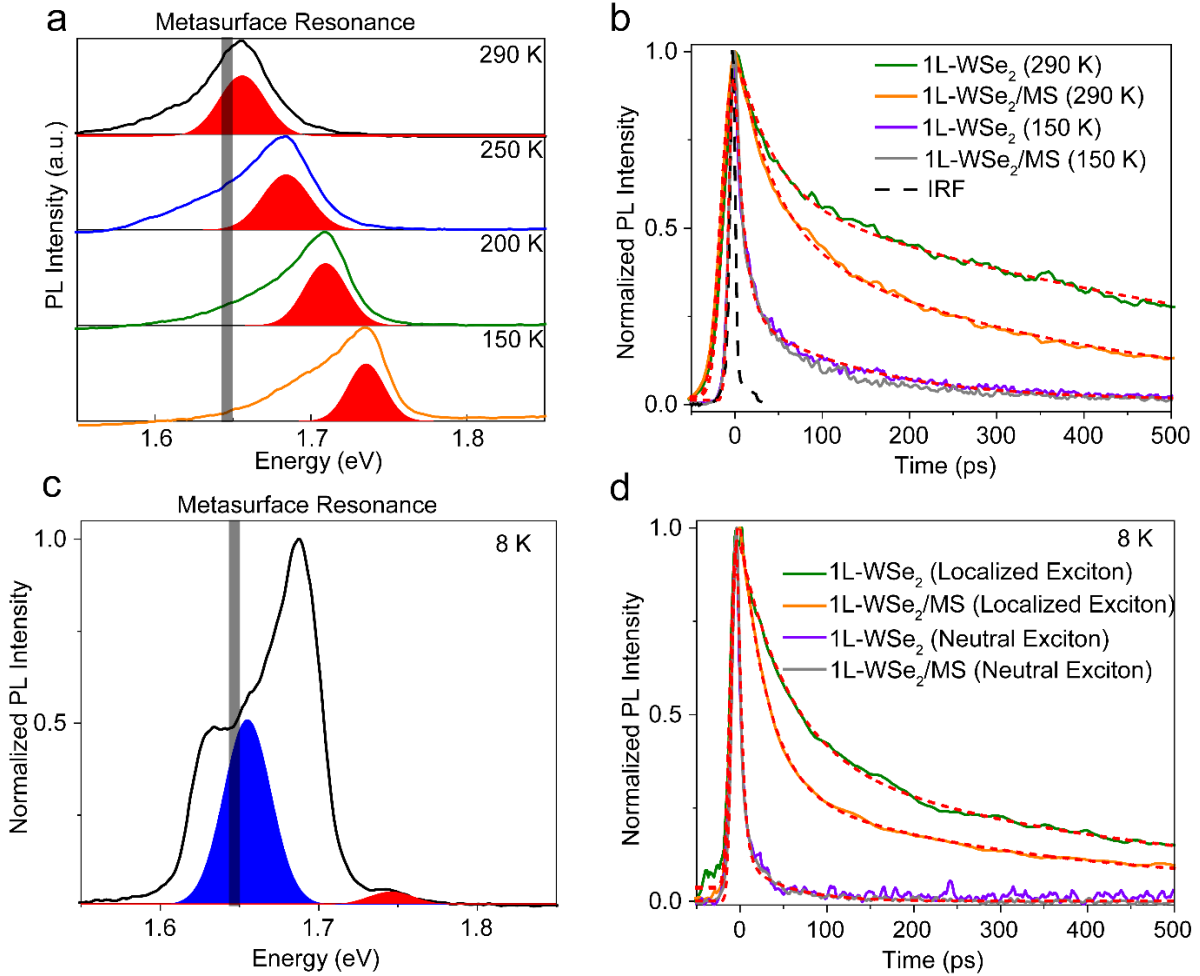

**Figure 3.** Tailoring exciton dynamics by controlling the Purcell effect through spectral overlap. (a) Normalized temperature-dependent PL spectra from 1L-WSe$_2$. The grey shaded area indicates the toroidal resonance of the metasurface. The red shaded area indicates the neutral exciton resonance. (b) Neutral exciton dynamics for bare 1L-WSe$_2$ and the 1L-WSe$_2$/MS heterostructure at T = 290 and 150 K with an exciton density of $3.3 \times 10^{11}$ cm$^{-2}$. The red dashed lines represent the fits using a bi-exponential function convoluted with a Gaussian function. The black dashed line is the instrument response function (IRF), showing a temporal resolution of ~ 5 ps. (c) PL spectra from the 1L-WSe$_2$/MS heterostructure at T = 8 K. The blue and red shaded areas indicate localized (1.65 eV) and neutral excitons (1.74 eV). (d) PL dynamics probed at the localized exciton and neutral exciton peaks for 1L-WSe$_2$ and 1L-WSe$_2$/MS at 8 K with a low exciton density ($3.3 \times 10^{11}$ cm$^{-2}$).

**Suppressing exciton-exciton annihilation by enhancing light-matter interactions**. We then built upon the results of the previous section to explore how density-dependent exciton dynamics (i.e., in the nonlinear regime) are modulated by the enhanced light-matter interactions in the 1L-WSe$_2$/MS. Figure 4a plots the PL dynamics at the neutral exciton resonance for bare 1L-WSe$_2$ as a function of exciton density at T = 290 K, showing that exciton dynamics become significantly faster at higher densities, likely due to exciton-exciton annihilation[8,32,45]. Exciton-exciton annihilation typically occurs at high exciton densities when two excitons interact with each other and one exciton recombines non-radiatively to the ground state by transferring its energy to the second exciton[46]. This process is enhanced due to the reduced dimensionality in TMD monolayers, which leads to strong many-body exciton interactions that in turn result in rapid exciton-exciton annihilation[8,9]. However, for the 1L-WSe$_2$/MS heterostructure, the neutral exciton PL decays significantly slower than that from bare 1L-WSe$_2$ at high exciton densities (Figure 4b), indicating a remarkably reduced exciton-exciton annihilation rate. We use the following rate equation, including a bimolecular annihilation term and exciton decay term, to describe the density-dependent PL dynamics[9,10,47]:

$$\frac{dN}{dt} = -\frac{N}{\tau} - k_A N^2 \qquad (2)$$

where $N$ is the exciton population, $k_A$ is the annihilation rate constant and $\tau$ is the exciton lifetime. The solution of the above equation can be written as:

$$\frac{1}{N(t)} = \left(\frac{1}{N(0)} + k_A \tau\right) \exp\left(\frac{t}{\tau}\right) - k_A \tau \qquad (3)$$

where $N(0)$ is the initial exciton density.

Since exciton-exciton annihilation is strongly exciton-density dependent, it has little effect on the dynamics at the lowest measured exciton density ($3.3 \times 10^{11}$ cm$^{-2}$). However, as shown in Figures 4c and 4d, this relatively simple annihilation model provides a good fit to the PL dynamics at higher exciton densities for both bare 1L-WSe$_2$ and the 1L-WSe$_2$/MS heterostructure. Considering all sets of data, we obtained average annihilation rate constants ($k_A$) of $1.1 \times 10^{-2}$ and $6.8 \times 10^{-3}$ cm$^2$/s for 1L-WSe$_2$ and 1L-WSe$_2$/MS, respectively, revealing that the exciton annihilation rate in the 1L-WSe$_2$/MS is reduced by a factor of 1.6 compared to that of bare 1L-WSe$_2$. This significant suppression of exciton-exciton annihilation is likely due to the Purcell effect in the 1L-WSe$_2$/MS, causing carriers to recombine radiatively before non-radiative processes like annihilation can occur. However, the exciton-exciton annihilation rate constant does not directly determine the PL lifetime. We can show this by qualitatively estimating the annihilation rate change between 1L-WSe$_2$/MS and bare 1L-WSe$_2$. We plotted the PL dynamics in bare 1L-WSe$_2$ and 1L-WSe$_2$/MS at 290 K for a low-density exciton population ($3.3 \times 10^{11}$ cm$^{-2}$), in which exciton-exciton annihilation is negligible, and a high-density case ($4.4 \times 10^{12}$ cm$^{-2}$), in which exciton-exciton annihilation dominates exciton recombination (Figure S6). The shaded area between low-density and high-density curves represents the total loss of exciton population due to exciton-exciton annihilation. Thus, we estimate that the exciton population in the 1L-WSe$_2$/MS is reduced by a factor of 2.8 compared with bare 1L-WSe$_2$; this agrees well with the ~2x reduction in the annihilation rate. This supports the idea that the Purcell effect-induced reduction in the exciton lifetime leads to a significant reduction of exciton-exciton annihilation upon integrating 1L-WSe$_2$ with the dielectric metasurface. Similar results have been observed previously in colloidal CdSe/CdS quantum dots coupled with plasmonic antennas[48]. The interplay between exciton-exciton annihilation and the Purcell effect is further demonstrated in Figure 4e. At a low exciton density ($3.3 \times 10^{11}$ cm$^{-2}$), the Purcell effect leads to faster exciton recombination in the 1L-WSe$_2$/MS as compared to bare 1L-WSe$_2$, due to the enhanced radiative decay rate and minimal exciton-exciton annihilation. However, as the exciton density increases, exciton-exciton annihilation begins to dominate the dynamics. Due to the significantly reduced annihilation rate, excitons live longer in the 1L-WSe$_2$/MS (188 ps) than bare 1L-WSe$_2$ (146 ps) at a density of $4.4 \times 10^{12}$ cm$^{-2}$.

Figure 4f plots the exciton-exciton annihilation rate constants as a function of temperature for both bare 1L-WSe$_2$ and the 1L-WSe$_2$/MS heterostructure, showing that they both display significantly faster annihilation at lower temperatures. Previous measurements have shown that exciton diffusion is faster at low temperatures, which likely increases the probability for exciton annihilation[32,49]. Furthermore, exciton-exciton annihilation is suppressed the most at 290 K, when the exciton resonance has the largest overlap with the toroidal resonance. As the temperature decreases, the spectral overlap between the neutral exciton resonance and toroidal resonance is gradually reduced, leading to a smaller difference in the annihilation rates for bare 1L-WSe$_2$ vs. 1L-WSe$_2$/MS (Figure 4f, inset). Finally, at 8 K, the metasurface resonance now overlaps well with the localized exciton resonance, leading to a significant reduction of the exciton-exciton annihilation rate in the 1L-WSe$_2$/MS ($7.4 \times 10^{-3}$ cm$^2$/s) compared with bare 1L-WSe$_2$ ($1.6 \times 10^{-2}$ cm$^2$/s) for the localized exciton (Figure S7). This further demonstrates the central role of spectral overlap in modulating exciton dynamics and annihilation, as well as our ability to control these processes via temperature tuning.

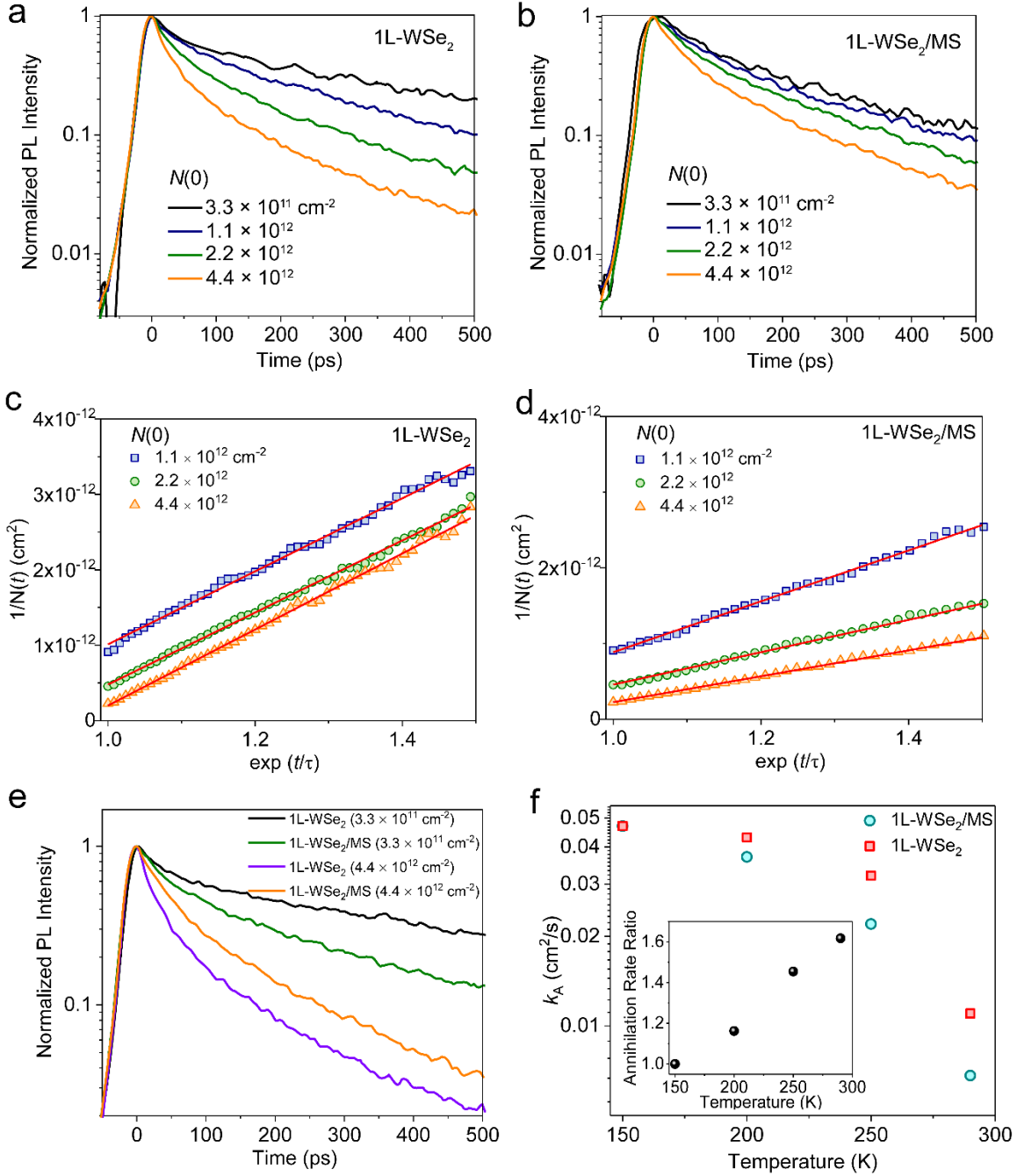

**Figure 4.** Suppressed exciton-exciton annihilation in the 1L-WSe$_2$/MS by enhanced light-matter interactions. (a, b) Exciton-density-dependent PL dynamics probed at the neutral exciton resonance in bare 1L-WSe$_2$ and 1L-WSe$_2$/MS at 290 K, respectively, for different values of $N$(0). (c, d) PL dynamics from (a) and (b), fit using the exciton-exciton annihilation model (Eq. (3)). (e) Selected PL dynamics for bare 1L-WSe$_2$ and 1L-WSe$_2$/MS at exciton densities of $3.3 \times 10^{11}$ and $4.4 \times 10^{12}$ cm$^{-2}$. (f) Temperature-dependent annihilation rate ($k_A$) for neutral excitons in 1L-WSe$_2$/MS and bare 1L-WSe$_2$. The inset depicts the annihilation rate ratio for bare and coupled samples.

## CONCLUSION

In summary, we have demonstrated enhanced light-matter interactions and control over exciton dynamics in 1L-WSe$_2$ when coupled with a toroidal dielectric metasurface. PL emission from the neutral exciton is enhanced over one order of magnitude in the 1L-WSe$_2$/MS heterostructure. Furthermore, by appropriately designing the metasurface, we can control exciton dynamics and annihilation through the temperature-dependent spectral overlap of exciton and toroidal resonances, allowing us to selectively enhance the Purcell effect and reduce exciton-exciton annihilation for different excitonic modes. Our results provide direct experimental evidence for strong dynamic light-matter interactions in single-layer TMDs when coupled with toroidal dielectric metasurfaces, with important implications for photonics applications including ultrafast sensors, light modulators, and quantum emitters.

## MATERIALS AND METHODS

**Sample preparation.** Atomic layer deposition was used to deposit an amorphous TiO$_2$ film on a quartz substrate, at 90 °C with a rate of 0.65 Å/cycle. After the deposition, electron beam resist ZEP520a was spin-coated and prebaked at 180 °C for 5 min. H$_2$O discharge solution was spin-coated on top of the resist to prevent possible charging in the following lithography step. The cuboid patterns were patterned with electron-beam lithography using an e-beam writer (JEOL), followed by development of the resist in N-amyl acetate for 1 min. A total of 15 nm of chromium metal was deposited and was lifted off in Remover-PG solution to serve as a hard mask. Fluorine and argon-based reactive ion etching etched out TiO$_2$ while leaving a chromium layer intact, transferring the patterns to the films. The chromium hard mask was later removed with CR-7 wet-etchant to finalize the sample.

**Transmission and reflection measurements.** To measure transmission and reflection from the metasurface and 1L-WSe$_2$, we used the white light from a lamp, sent it through a polarizer and weakly focused it onto the sample using an 50X objective with numerical aperture of 0.55. Transmitted and reflected beams were sent to a spectrometer containing an Si array detector. Transmission through the sample was divided by that through the bare substrate to obtain a normalized spectrum.

**Ultrafast PL spectroscopy.** Steady-state PL and time-resolved PL measurements were performed by employing a home-built confocal PL setup. An ultrafast Ti:sapphire laser (Coherent Chameleon) operating at a repetition rate of 80 MHz, together with a second harmonic generation module, was used to produce an excitation beam at a photon energy of 2.48 eV. The laser beam was focused by a 100X objective with a numerical aperture of 0.7 to a diffraction-limited beam spot (~ 400 nm, Figure S8). The PL emission was collected with the same objective and detected by a streak camera (Hamamatsu). The time resolution of this setup is ~ 5 ps. Temperature-dependent PL measurements were carried out in a closed-cycle cryostat.

## AUTHOR CONTRIBUTIONS

#L.Y. and J.J. contributed equally to this work. R.P.P. and I.B. designed the experiments. L.Y. and J.J. carried out optical measurements. J.J., K.W. C. K. and E.S.Y fabricated the sample. M.G., D. R. and J.C.H grew and provided the bulk $WSe_2$ crystal. L.Y., J.J., I.B. and R.P.P. analyzed the experimental data and wrote the manuscript with input from all authors.

**Notes**

The authors declare no competing interests.

## ACKNOWLEDGEMENTS

This work was supported by the U.S. Department of Energy, Office of Basic Energy Sciences, Division of Materials Sciences and Engineering and performed, in part, at the Center for Integrated Nanotechnologies, an Office of Science User Facility operated for the U.S. Department of Energy (DOE) Office of Science. L.Y. and R.P.P. acknowledge financial support from the LANL LDRD Program. Sandia National Laboratories is a multi-mission laboratory managed and operated by National Technology and Engineering Solutions of Sandia, LLC, a wholly owned subsidiary of Honeywell International, Inc., for the U.S. Department of Energy's National Nuclear Security Administration under contract DE-NA0003525. Los Alamos National Laboratory, an affirmative action equal opportunity employer, is managed by Triad National Security, LLC for the U.S. Department of Energy's NNSA, under Contract No. 89233218CNA000001. J.J acknowledges the support from a National Research Foundation of Korea (NRF) grant funded by the Korean Government (MSIT: NRF-2021R1C1C1010660), and a 2020 Research Grant from Kangwon National University. Growth of $WSe_2$ (E.S.Y, M.G., D. R. and J.C.H) was supported at Columbia through the NSF MRSEC program (DMR-2011738)This paper describes objective technical

**REFERENCES**

(1) Novoselov, K. S.; Jiang, D.; Schedin, F.; Booth, T. J.; Khotkevich, V. V.; Morozov, S. V.; Geim, A. K. Two-dimensional atomic crystals. *Proc. Natl. Acad. Sci. U.S.A.* **2005**, 10451.
(2) Wang, Q. H.; Kalantar-Zadeh, K.; Kis, A.; Coleman, J. N.; Strano, M. S. Electronics and optoelectronics of two-dimensional transition metal dichalcogenides. *Nat. Nanotechnol.* **2012**, 699-712.
(3) Mak, K. F.; Lee, C.; Hone, J.; Shan, J.; Heinz, T. F. Atomically Thin $MoS_2$: A New Direct-Gap Semiconductor. *Phys. Rev. Lett.* **2010**, 136805.
(4) Manzeli, S.; Ovchinnikov, D.; Pasquier, D.; Yazyev, O. V.; Kis, A. 2D transition metal dichalcogenides. *Nat. Rev. Mater.* **2017**, 17033.
(5) Chernikov, A.; Berkelbach, T. C.; Hill, H. M.; Rigosi, A.; Li, Y.; Aslan, O. B.; Reichman, D. R.; Hybertsen, M. S.; Heinz, T. F. Exciton Binding Energy and Nonhydrogenic Rydberg Series in Monolayer $WS_2$. *Phys. Rev. Lett.* **2014**, 076802.
(6) Bernardi, M.; Palummo, M.; Grossman, J. C. Extraordinary Sunlight Absorption and One Nanometer Thick Photovoltaics Using Two-Dimensional Monolayer Materials. *Nano Lett.* **2013**, 3664-3670.
(7) Jin, C.; Kim, J.; Wu, K.; Chen, B.; Barnard, E. S.; Suh, J.; Shi, Z.; Drapcho, S. G.; Wu, J.; Schuck, P. J.; Tongay, S.; Wang, F. On Optical Dipole Moment and Radiative Recombination Lifetime of Excitons in $WSe_2$. *Adv. Funct. Mater.* **2017**, 1601741.
(8) Sun, D.; Rao, Y.; Reider, G. A.; Chen, G.; You, Y.; Brézin, L.; Harutyunyan, A. R.; Heinz, T. F. Observation of Rapid Exciton–Exciton Annihilation in Monolayer Molybdenum Disulfide. *Nano Lett.* **2014**, 5625-5629.
(9) Yuan, L.; Huang, L. Exciton dynamics and annihilation in $WS_2$ 2D semiconductors. *Nanoscale* **2015**, 7402-7408.
(10) Kumar, N.; Cui, Q.; Ceballos, F.; He, D.; Wang, Y.; Zhao, H. Exciton-exciton annihilation in $MoSe_2$ monolayers. *Phys. Rev. B: Condens. Matter Mater. Phys.* **2014**, 125427.
(11) Cunningham, P. D.; McCreary, K. M.; Jonker, B. T. Auger Recombination in Chemical Vapor Deposition-Grown Monolayer $WS_2$. *J. Phys. Chem. Lett.* **2016**, 5242-5246.
(12) Ginn, J. C.; Brener, I.; Peters, D. W.; Wendt, J. R.; Stevens, J. O.; Hines, P. F.; Basilio, L. I.; Warne, L. K.; Ihlefeld, J. F.; Clem, P. G.; Sinclair, M. B. Realizing Optical Magnetism from Dielectric Metamaterials. *Phys. Rev. Lett.* **2012**, 097402.
(13) Kuznetsov, A. I.; Miroshnichenko, A. E.; Brongersma, M. L.; Kivshar, Y. S.; Luk'yanchuk, B. Optically resonant dielectric nanostructures. *Science* **2016**, aag2472.
(14) Cambiasso, J.; Grinblat, G.; Li, Y.; Rakovich, A.; Cortés, E.; Maier, S. A. Bridging the Gap between Dielectric Nanophotonics and the Visible Regime with Effectively Lossless Gallium Phosphide Antennas. *Nano Lett.* **2017**, 1219-1225.
(15) Sortino, L.; Zotev, P. G.; Mignuzzi, S.; Cambiasso, J.; Schmidt, D.; Genco, A.; Aßmann, M.; Bayer, M.; Maier, S. A.; Sapienza, R.; Tartakovskii, A. I. Enhanced light-matter interaction in an atomically thin semiconductor coupled with dielectric nano-antennas. *Nat. Commun.* **2019**, 5119.
(16) Jeong, P. A.; Goldflam, M. D.; Campione, S.; Briscoe, J. L.; Vabishchevich, P. P.; Nogan, J.; Sinclair, M. B.; Luk, T. S.; Brener, I. High Quality Factor Toroidal Resonances in Dielectric Metasurfaces. *ACS Photonics* **2020**, 1699-1707.
(17) Staude, I.; Pertsch, T.; Kivshar, Y. S. All-Dielectric Resonant Meta-Optics Lightens up. *ACS Photonics* **2019**, 802-814.
(18) Ma, X.; James, A. R.; Hartmann, N. F.; Baldwin, J. K.; Dominguez, J.; Sinclair, M. B.; Luk, T. S.; Wolf, O.; Liu, S.; Doorn, S. K.; Htoon, H.; Brener, I. Solitary Oxygen Dopant Emission from Carbon Nanotubes Modified by Dielectric Metasurfaces. *ACS Nano* **2017**, 6431-6439.
(19) Bucher, T.; Vaskin, A.; Mupparapu, R.; Löchner, F. J. F.; George, A.; Chong, K. E.; Fasold, S.; Neumann, C.; Choi, D.-Y.; Eilenberger, F.; Setzpfandt, F.; Kivshar, Y. S.; Pertsch, T.; Turchanin, A.;

Staude, I. Tailoring Photoluminescence from MoS$_2$ Monolayers by Mie-Resonant Metasurfaces. *ACS Photonics* **2019**, 1002-1009.

(20) Brongersma, M. L. The road to atomically thin metasurface optics. *Nanophotonics* **2021**, 643-654.

(21) Liu, S.; Vaskin, A.; Addamane, S.; Leung, B.; Tsai, M.-C.; Yang, Y.; Vabishchevich, P. P.; Keeler, G. A.; Wang, G.; He, X.; Kim, Y.; Hartmann, N. F.; Htoon, H.; Doorn, S. K.; Zilk, M.; Pertsch, T.; Balakrishnan, G.; Sinclair, M. B.; Staude, I.; Brener, I. Light-Emitting Metasurfaces: Simultaneous Control of Spontaneous Emission and Far-Field Radiation. *Nano Lett.* **2018**, 6906-6914.

(22) Siday, T.; Vabishchevich, P. P.; Hale, L.; Harris, C. T.; Luk, T. S.; Reno, J. L.; Brener, I.; Mitrofanov, O. Terahertz Detection with Perfectly-Absorbing Photoconductive Metasurface. *Nano Lett.* **2019**, 2888-2896.

(23) Zhao, W.; Ghorannevis, Z.; Chu, L.; Toh, M.; Kloc, C.; Tan, P.-H.; Eda, G. Evolution of Electronic Structure in Atomically Thin Sheets of WS$_2$ and WSe$_2$. *ACS Nano* **2013**, 791-797.

(24) Kozawa, D.; Kumar, R.; Carvalho, A.; Kumar Amara, K.; Zhao, W.; Wang, S.; Toh, M.; Ribeiro, R. M.; Castro Neto, A. H.; Matsuda, K.; Eda, G. Photocarrier relaxation pathway in two-dimensional semiconducting transition metal dichalcogenides. *Nat. Commun.* **2014**, 4543.

(25) Wang, Z.; Dong, Z.; Gu, Y.; Chang, Y.-H.; Zhang, L.; Li, L.-J.; Zhao, W.; Eda, G.; Zhang, W.; Grinblat, G.; Maier, S. A.; Yang, J. K. W.; Qiu, C.-W.; Wee, A. T. S. Giant photoluminescence enhancement in tungsten-diselenide–gold plasmonic hybrid structures. *Nat. Commun.* **2016**, 11283.

(26) Huang, J.; Akselrod, G. M.; Ming, T.; Kong, J.; Mikkelsen, M. H. Tailored Emission Spectrum of 2D Semiconductors Using Plasmonic Nanocavities. *ACS Photonics* **2018**, 552-558.

(27) Chen, Y.; Miao, S.; Wang, T.; Zhong, D.; Saxena, A.; Chow, C.; Whitehead, J.; Gerace, D.; Xu, X.; Shi, S.-F.; Majumdar, A. Metasurface Integrated Monolayer Exciton Polariton. *Nano Lett.* **2020**, 5292-5300.

(28) Liu, C.-H.; Clark, G.; Fryett, T.; Wu, S.; Zheng, J.; Hatami, F.; Xu, X.; Majumdar, A. Nanocavity Integrated van der Waals Heterostructure Light-Emitting Tunneling Diode. *Nano Lett.* **2017**, 200-205.

(29) Borghardt, S.; Tu, J.-S.; Winkler, F.; Schubert, J.; Zander, W.; Leosson, K.; Kardynał, B. E. Engineering of optical and electronic band gaps in transition metal dichalcogenide monolayers through external dielectric screening. *Phys. Rev. Mater.* **2017**, 054001.

(30) Purcell, E. M. Spontaneous Emission Probabilities at Radio Frequencies. In *Confined Electrons and Photons: New Physics and Applications*; Burstein, E.; Weisbuch, C., Eds.; Springer US: Boston, MA, 1995; pp 839-839.

(31) You, Y.; Zhang, X.-X.; Berkelbach, T. C.; Hybertsen, M. S.; Reichman, D. R.; Heinz, T. F. Observation of biexcitons in monolayer WSe$_2$. *Nat. Phys.* **2015**, 477-481.

(32) Mouri, S.; Miyauchi, Y.; Toh, M.; Zhao, W.; Eda, G.; Matsuda, K. Nonlinear photoluminescence in atomically thin layered WSe$_2$ arising from diffusion-assisted exciton-exciton annihilation. *Phys. Rev. B: Condens. Matter Mater. Phys.* **2014**, 155449.

(33) Varshni, Y. P. Temperature dependence of the energy gap in semiconductors. *Physica* **1967**, 149-154

(34) Huang, J.; Hoang, T. B.; Mikkelsen, M. H. Probing the origin of excitonic states in monolayer WSe$_2$. *Sci. Rep.* **2016**, 22414.

(35) Amani, M.; Taheri, P.; Addou, R.; Ahn, G. H.; Kiriya, D.; Lien, D.-H.; Ager, J. W.; Wallace, R. M.; Javey, A. Recombination Kinetics and Effects of Superacid Treatment in Sulfur- and Selenium-Based Transition Metal Dichalcogenides. *Nano Lett.* **2016**, 2786-2791.

(36) Selig, M.; Berghäuser, G.; Raja, A.; Nagler, P.; Schüller, C.; Heinz, T. F.; Korn, T.; Chernikov, A.; Malic, E.; Knorr, A. Excitonic linewidth and coherence lifetime in monolayer transition metal dichalcogenides. *Nat. Commun.* **2016**, 13279.

(37) Moody, G.; Kavir Dass, C.; Hao, K.; Chen, C.-H.; Li, L.-J.; Singh, A.; Tran, K.; Clark, G.; Xu, X.; Berghäuser, G.; Malic, E.; Knorr, A.; Li, X. Intrinsic homogeneous linewidth and broadening mechanisms of excitons in monolayer transition metal dichalcogenides. *Nat. Commun.* **2015**, 8315.

(38) Zhang, X.-X.; You, Y.; Zhao, S. Y. F.; Heinz, T. F. Experimental Evidence for Dark Excitons in Monolayer WSe$_2$. *Phys. Rev. Lett.* **2015**, 257403.

(39) Li, Z.; Wang, T.; Lu, Z.; Khatoniar, M.; Lian, Z.; Meng, Y.; Blei, M.; Taniguchi, T.; Watanabe, K.; McGill, S. A.; Tongay, S.; Menon, V. M.; Smirnov, D.; Shi, S.-F. Direct Observation of Gate-Tunable Dark Trions in Monolayer $WSe_2$. *Nano Lett.* **2019**, 6886-6893.

(40) Li, Z.; Wang, T.; Jin, C.; Lu, Z.; Lian, Z.; Meng, Y.; Blei, M.; Gao, M.; Taniguchi, T.; Watanabe, K.; Ren, T.; Cao, T.; Tongay, S.; Smirnov, D.; Zhang, L.; Shi, S.-F. Momentum-Dark Intervalley Exciton in Monolayer Tungsten Diselenide Brightened via Chiral Phonon. *ACS Nano* **2019**, 14107-14113.

(41) Li, Z.; Wang, T.; Jin, C.; Lu, Z.; Lian, Z.; Meng, Y.; Blei, M.; Gao, S.; Taniguchi, T.; Watanabe, K.; Ren, T.; Tongay, S.; Yang, L.; Smirnov, D.; Cao, T.; Shi, S.-F. Emerging photoluminescence from the dark-exciton phonon replica in monolayer $WSe_2$. *Nat. Commun.* **2019**, 2469.

(42) Wang, G.; Bouet, L.; Lagarde, D.; Vidal, M.; Balocchi, A.; Amand, T.; Marie, X.; Urbaszek, B. Valley dynamics probed through charged and neutral exciton emission in monolayer $WSe_2$. *Phys. Rev. B: Condens. Matter Mater. Phys.* **2014**, 075413.

(43) Wagner, J.; Kuhn, H.; Bernhardt, R.; Zhu, J.; van Loosdrecht, P. H. M. Trap induced long exciton intervalley scattering and population lifetime in monolayer $WSe_2$. *2D Mater.* **2021**, 035018.

(44) Luo, Y.; Shepard, G. D.; Ardelean, J. V.; Rhodes, D. A.; Kim, B.; Barmak, K.; Hone, J. C.; Strauf, S. Deterministic coupling of site-controlled quantum emitters in monolayer $WSe_2$ to plasmonic nanocavities. *Nat. Nanotechnol.* **2018**, 1137-1142.

(45) Yuan, L.; Wang, T.; Zhu, T.; Zhou, M.; Huang, L. Exciton Dynamics, Transport, and Annihilation in Atomically Thin Two-Dimensional Semiconductors. *J. Phys. Chem. Lett.* **2017**, 3371-3379.

(46) Htoon, H.; Hollingsworth, J. A.; Dickerson, R.; Klimov, V. I. Effect of Zero- to One-Dimensional Transformation on Multiparticle Auger Recombination in Semiconductor Quantum Rods. *Phys. Rev. Lett.* **2003**, 227401.

(47) Shaw, P. E.; Ruseckas, A.; Samuel, I. D. W. Exciton Diffusion Measurements in Poly(3-hexylthiophene). *Adv. Mater.* **2008**, 3516-3520.

(48) Dhawan, A. R.; Belacel, C.; Esparza-Villa, J. U.; Nasilowski, M.; Wang, Z.; Schwob, C.; Hugonin, J.-P.; Coolen, L.; Dubertret, B.; Senellart, P.; Maître, A. Extreme multiexciton emission from deterministically assembled single-emitter subwavelength plasmonic patch antennas. *Light Sci. Appl.* **2020**, 33.

(49) Cadiz, F.; Robert, C.; Courtade, E.; Manca, M.; Martinelli, L.; Taniguchi, T.; Watanabe, K.; Amand, T.; Rowe, A. C. H.; Paget, D.; Urbaszek, B.; Marie, X. Exciton diffusion in $WSe_2$ monolayers embedded in a van der Waals heterostructure. *Appl. Phys. Lett.* **2018**, 152106.